\documentstyle[12pt,epsfig]{article}

\catcode`\@=11
\long\def\@makefntext#1{
\protect\noindent \hbox to 3.2pt {\hskip-.9pt
$^{{\ninerm\@thefnmark}}$\hfil}#1\hfill}                %CAN BE USED

\def\@makefnmark{\hbox to 0pt{$^{\@thefnmark}$\hss}}  %ORIGINAL

\def\ps@myheadings{\let\@mkboth\@gobbletwo
\def\@oddhead{\hbox{}
\rightmark\hfil\ninerm\thepage}
\def\@oddfoot{}\def\@evenhead{\ninerm\thepage\hfil
\leftmark\hbox{}}\def\@evenfoot{}
\def\sectionmark##1{}\def\subsectionmark##1{}}

\renewcommand{\thefootnote}{\fnsymbol{footnote}}

\def\sectionc{\@startsection {section}{1}{\z@}{-3.5ex plus -1ex minus
    -.2ex}{2.3ex plus .2ex}{\bf }}
\def\subsectionc{\@startsection{subsection}{2}{\z@}{-3.25ex plus -1ex minus
   -.2ex}{1.5ex plus .2ex}{\it }}
\renewcommand{\section}[1]{\sectionc{#1}\hspace*{\parindent}}
\renewcommand{\subsection}[1]{\subsectionc{#1}\hspace*{\parindent}}
\newcounter{appendixc}
\newcounter{subappendixc}[appendixc]
\newcounter{subsubappendixc}[subappendixc]

\renewcommand{\appendix}[1] {\vspace*{0.6cm}
        \refstepcounter{appendixc}
        \setcounter{figure}{0}
        \setcounter{table}{0}
        \setcounter{equation}{0}
        \renewcommand{\thefigure}{\Alph{appendixc}.\arabic{figure}}
        \renewcommand{\thetable}{\Alph{appendixc}.\arabic{table}}
        \renewcommand{\theappendixc}{\Alph{appendixc}}
        \renewcommand{\theequation}{\Alph{appendixc}.\arabic{equation}}
        \noindent{\bf Appendix \theappendixc #1}\par\vspace*{0.4cm}}

\def\abstracts#1{{

\centering{\begin{minipage}{13.2truecm}\footnotesize\baselineskip=13pt\noindent
        \parindent=0pt #1
        \end{minipage}}\par}}

\renewenvironment{thebibliography}[1]
        {\begin{list}{\arabic{enumi}.}
        {\usecounter{enumi}\setlength{\parsep}{0pt}
\setlength{\leftmargin 0.75cm}{\rightmargin 0pt}
         \setlength{\itemsep}{0pt} \settowidth
        {\labelwidth}{#1.}\sloppy}}{\end{list}}

\newcounter{itemlistc}
\newcounter{romanlistc}
\newcounter{alphlistc}
\newcounter{arabiclistc}

\newcommand{\fcaption}[1]{
        \refstepcounter{figure}
        \setbox\@tempboxa = \hbox{\footnotesize Figure~\thefigure. #1}
        \ifdim \wd\@tempboxa > 6in
           {\begin{center}
        \parbox{6in}{\footnotesize\baselineskip=13pt Figure~\thefigure. #1}
            \end{center}}
        \else
             {\begin{center}
             {\footnotesize Figure~\thefigure. #1}
              \end{center}}
        \fi}

\newcommand{\tcaption}[1]{
        \refstepcounter{table}
        \setbox\@tempboxa = \hbox{\footnotesize Table~\thetable. #1}
        \ifdim \wd\@tempboxa > 6in
           {\begin{center}
        \parbox{6in}{\footnotesize\baselineskip=13pt Table~\thetable. #1}
            \end{center}}
        \else
             {\begin{center}
             {\footnotesize Table~\thetable. #1}
              \end{center}}
        \fi}

\def\@citex[#1]#2{\if@filesw\immediate\write\@auxout
        {\string\citation{#2}}\fi
\def\@citea{}\@cite{\@for\@citeb:=#2\do
        {\@citea\def\@citea{,}\@ifundefined
        {b@\@citeb}{{\bf ?}\@warning
        {Citation `\@citeb' on page \thepage \space undefined}}
        {\csname b@\@citeb\endcsname}}}{#1}}

\newif\if@cghi
\def\cite{\@cghitrue\@ifnextchar [{\@tempswatrue
        \@citex}{\@tempswafalse\@citex[]}}
\def\citelow{\@cghifalse\@ifnextchar [{\@tempswatrue
        \@citex}{\@tempswafalse\@citex[]}}
\def\@cite#1#2{{$\null^{#1}$\if@tempswa\typeout
        {IJCGA warning: optional citation argument
        ignored: `#2'} \fi}}

 1
 1
 1

\font\ninerm=cmr9

\begin{document}
% my definitions
\def\bra#1{{\langle #1{\left| \right.}}}
\def\ket#1{{{\left.\right|} #1\rangle}}
\def\bfgreek#1{ \mbox{\boldmath$#1$}}
\def\lesssim{\mathrel{\mathpalette\vereq<}}
\def\vereq#1#2{\lower3pt\vbox{\baselineskip1.5pt \lineskip1.5pt
\ialign{$\m@th#1\hfill##\hfil$\crcr#2\crcr\sim\crcr}}}

\def\gtrsim{\mathrel{\mathpalette\vereq>}}

\def\alt{\lesssim}
\def\agt{\gtrsim}

%\centerline{\normalsize\bf WORLD SCIENTIFIC PUBLISHING COMPANY GUIDELINES FOR}
%\baselineskip=15pt
%\centerline{\normalsize\bf TYPESETTING A CAMERA-READY MANUSCRIPT}
%\centerline{\footnotesize\sf (For subsequent 20\% photoreduction to
%17 $\times$ 12 cm text area)\footnote{The \LaTeX\ source file 
%for this document
%may be used as a template for your article.}}
\centerline{\normalsize\bf Delta electroproduction in a chiral bag model approach}
%Galilean invariant cloudy bag model}
%A chiral bag model approach to delta electroproduction}
%\vfill
\vspace*{0.6cm}
\centerline{\footnotesize D. H. Lu, A. W. Thomas, and  A. G. Williams}
\baselineskip=13pt
\centerline{\footnotesize\it Department of Physics and Mathematical Physics,}
\baselineskip=13pt
\centerline{\footnotesize\it University of Adelaide, Australia 5005}
\centerline{\footnotesize\it        and}
\centerline{\footnotesize\it Institute for Theoretical Physics,
	University of Adelaide, Australia 5005}
\baselineskip=13pt
\centerline{\footnotesize E-mail: dlu,athomas,awilliam@physics.adelaide.edu.au}
\vspace*{0.3cm}
%\centerline{\footnotesize and}
%\vspace*{0.3cm}
%\centerline{\footnotesize SECOND AUTHOR'S NAME}
%\baselineskip=13pt
%\centerline{\footnotesize\it Group, Company, Address, City, State ZIP/Zone,
%Country}

%\vfill
\vspace*{0.6cm}
\abstracts{We study the $\gamma^* N \rightarrow\Delta$
transition amplitudes in a recoil corrected cloudy bag model approach.
A modified  Peierls-Thouless projection method is used to construct
the Galilean invariant baryon states. The pionic contribution is found to be
significant. The effect of the recoil correction is to reduce the 
magnitude of the transition amplitudes at small momentum transfer and 
to enhance them at modest momentum transfers.}

%{Helicity amplitudes for the $\gamma^* N \rightarrow\Delta$ 
%transition are calculated using the cloudy bag model.
%A correction for center-of-mass motion is carried out
%using a modified  Peierls-Thouless projection method. This reduces the 
%magnitude of the transition amplitudes at small momentum transfer and 
%enhances them at modest momentum transfers.
%Our calculation shows that the pion cloud contributes substantially
%to the transition helicity amplitudes, with the final result giving
%reasonable agreement with the corresponding experimental values.}}

%\vspace*{0.6cm}
\normalsize\baselineskip=15pt
\setcounter{footnote}{0}
\renewcommand{\thefootnote}{\alph{footnote}}

\section{Introduction}\label{sec:general}
The nucleon-delta electromagnetic transition amplitude is
an outstanding example of the success of the quark model.
There have been many theoretical and experimental explorations 
of this transition process. 
In a naive quark model the $\gamma^* N\rightarrow\Delta$ 
transition occurs only 
by an M1 transition, while the E2 process is fully suppressed\cite{BECCHI65}.
In  more sophisticated models, quarks can interact through, for example,
 one-gluon 
exchange in addition to the confinement potential between them. Then it is 
possible for configuration mixing, involving the excitation of one quark to 
a $d$-state, to
generate a  small, but nonvanishing, E2 amplitude\cite{IK82}.
To extract the $\gamma^* N\rightarrow\Delta$ amplitude from experimental data 
is  not an easy task. 
There are some uncertainties in the subtraction of  background,
 and the results are somewhat model dependent\cite{DAVID91}. 
With the advent of the new generation of accelerators,
much more accurate measurements will be made. 
The anticipated high quality
data  should test various hadron models 
and help to build  more realistic ones.

The cloudy  bag model (CBM)\cite{CBM} improves the MIT bag model\cite{MIT} 
by introducing 
an elementary, perturbative pion field which 
couples to quarks in the bag in such a way 
that chiral symmetry is restored.
The  pion field significantly improves the predictions of the static 
 properties of baryons.
Previous calculations of delta photoproduction\cite{KE83,Bermuth86} 
in the cloudy bag model differ from the results presented here and 
neglected the recoil correction.
The baryon wave function is simply a direct product of individual
quark wave functions, similar to the  nuclear shell-model 
wave function (independent particle motion).
This type of wave function is not a momentum eigenstate although
the Hamiltonian commutes with the total momentum operator.
The matrix elements evaluated between such static states contain 
spurious center of mass 
motion which ought to be removed.
Early studies indicated that the  correction for spurious 
 center of mass motion is significant\cite{Wilets}.
It is expected to be most important in calculations where  relatively
large momentum transfers are involved.
There are several intuitively motivated prescriptions\cite{Wilets}
 for the correction of
center of mass motion, however, none of them are fully satisfactory.
In this work, we have chosen to use  the  Peierls-Thouless (PT)\cite{PT62} 
method to eliminate the center of mass motion, since it is the most
convenient for our purposes. 

In this paper, we calculate the nucleon-delta electromagnetic 
transition amplitudes
with respect to the virtual photon. The spurious center of mass
motion is corrected by using the PT projection method. 
As a first step, we assume exact $SU(6)$ symmetry for the quark structure 
of the baryons, so that all
quarks in the ground state of the  $N$ and $\Delta$ are in the $s$ state.
The notation of references\cite{TT83,THESIS} is followed.
We briefly review  the method to construct
 the PT wave function in Sec.~II.\@ \, 
The calculation of helicity amplitudes for
virtual photoproduction of the delta is performed in Sec.~III, 
and in Sec.~IV we present the numerical results. Finally in Sec.~V we
summarize our results.

\section{Galilean invariant baryon states}
We start with the chirally invariant Lagrangian density 
of the cloudy bag model\cite{CBM}
\begin{eqnarray}
        \protect{\cal L}
        &=&  (i\overline q \gamma^\mu \partial_\mu q - B)\theta_V 
        - {1\over 2}\overline q q \Delta_S \nonumber \\
        && + {1\over 2} (\partial_\mu \bfgreek{\pi})^2 
        - {1\over 2} m^2_\pi \bfgreek{\pi}^2
        - {i\over 2f} \overline q \gamma_5 \bfgreek{\tau} \cdot 
        \bfgreek{\pi} q \Delta_S, \label{L}
\end{eqnarray}
where $\theta_V$ is a step function which is one inside the bag volume V 
and vanishes  outside, and
$\Delta_S$ is  a surface delta function.
In a perturbative treatment of the pion field, the quark wave function is not 
affected by the pion field and is given by the MIT bag solution\cite{MIT}
\begin{equation}
q({\bf r}) =  \left ( 
\begin{array}{c} g(r) \\ i\bfgreek {\sigma} \cdot\hat{r} f(r) \end{array} \right
 )  \theta(R-r),
\end{equation}
where $R$ is the spherical bag radius. 
For the ground state of a massless quark 
$g(r) = N_s j_0(\omega_s r/R), f(r) = N_s j_1(\omega_s r/R) $, 
where  $\omega_s = 2.04$ and 
 $N_s^2 = \omega_s/8\pi R^3 j_0^2(\omega_s) (\omega_s -1)$.  

The  bare baryon is taken to be composed of three quarks with the spin-flavor  
wave function  given by $SU(6)$ symmetry. Naively the space
component is  the  direct product of three
quark wave functions in coordinate space
\begin{equation}
\Psi({\bf x}_1, {\bf x}_2, {\bf x}_3; {\bf x}) = 
q({\bf x}_1 - {\bf x}) q({\bf x}_2 - {\bf x}) q({\bf x}_3- {\bf x}).
\end{equation}
Here  ${\bf x}$ indicates the location of the bag center, while 
${\bf x}_1$, ${\bf x}_2$, and ${\bf x}_3$ specify the positions 
of the three quarks. Clearly this wavefunction 
does not have definite momentum and is 
not a momentum eigenstate. A momentum eigenstate 
of the  baryon can be constructed 
by making a linear superposition of the localized states, namely,
\begin{equation}
\Psi_{\rm{PY}}({\bf x}_1, {\bf x}_2, {\bf x}_3; {\bf p})
= N'(p) \int\! d^3{\bf x} e^{i {\bf p\cdot x}}
\Psi({\bf x}_1, {\bf x}_2, {\bf x}_3; {\bf x}),
\end{equation}
where the subscript PY stands for Peierls-Yoccoz projection\cite{PY57}, 
and $N'(p)$ is a momentum dependent normalization constant.
It can be shown that $\Psi_{\rm{PY}}({\bf p}) =
e^{i{\bf p\cdot x}_{\rm{cm}}} \Psi_{\rm{in}}({\bf p})$, where 
${\bf x}_{\rm{cm}} = ({\bf x}_1 + {\bf x}_2 + {\bf x}_3)/3$ 
is the the center of mass of the baryon
and $\Psi_{\rm{in}}({\bf p})$ is the appropriately defined intrinsic part of 
the wave function. 
Since $\Psi_{\rm{in}}({\bf p})$  still depends on the c.m. momentum, 
it violates  translational invariance. To overcome this problem, 
Peierls and Thouless (PT)\cite{PT62} proposed
to make another superposition of these momentum eigenstates, i.e.,
\begin{equation}
\Psi_{\rm{PT}}({\bf x}_1, {\bf x}_2, {\bf x}_3; {\bf p})
= N(p) \int\! d^3p' w({\bf p'}) e^{i({\bf p - p'}) \cdot {\bf x}_{\rm{cm}} }
\Psi_{\rm{PY}}({\bf x}_1, {\bf x}_2, {\bf x}_3; {\bf p'}).
\end{equation}
The weight function, $w(p')$,  should in fact be chosen to minimize
the total energy, but this would be quite complicated to implement here.
Instead, we choose $w({\bf p'}) = 1$ for simplicity
and convenience.  
Then integrations over  ${\bf x}$ and ${\bf p'}$ can be carried out easily.
This leads to a much simplified PT wave function,
\begin{equation}
\Psi_{\rm{PT}}({\bf x}_1, {\bf x}_2, {\bf x}_3; {\bf p}) = 
N_{\rm{PT}} e^{i{\bf p \cdot x}_{\rm{cm}}}
q({\bf x}_1 - {\bf x}_{\rm{cm}}) q({\bf x}_2 - {\bf x}_{\rm{cm}})
q({\bf x}_3 - {\bf x}_{\rm{cm}}), \label{PT}
\end{equation}
where the normalization factor, $N_{\rm{PT}}$, is given by the condition
\begin{equation}
\int\! d^3 x_1 d^3 x_2 d^3 x_3  
\Psi^\dagger_{\rm{PT}}({\bf x}_1, {\bf x}_2, {\bf x}_3; {\bf p'})
\Psi_{\rm{PT}}        ({\bf x}_1, {\bf x}_2, {\bf x}_3; {\bf p}) 
= (2\pi)^3 \delta^{(3)}(\bf p' - \bf p).
\end{equation}
This leads to 
\begin{equation}
N_{\rm{PT}} = \left[ 3\int\! d^3 r_1 d^3 r_2 \rho({\bf r}_1) \rho({\bf r}_2) 
\rho(-{\bf r}_1 - {\bf r}_2) \right ]^{-1/2},
\end{equation}
where $\rho({\bf r})
\equiv q^\dagger({\bf r}) q({\bf r}) = [g^2(r) + f^2(r)]\, \theta(R-r)$.
Notice that, for this simple version of the PT projection,
$N_{\rm{PT}}$ is a momentum independent constant and 
the wavefunction in Eq.~(\ref{PT}) is manifestly Galilean invariant. 

\section{The helicity amplitudes in the cloudy bag model}
From the CBM Lagrangian given in  Eq.~(\ref{L}), the conserved local  
electromagnetic current can be derived using 
 the principle of minimal coupling 
$\partial_\mu \rightarrow \partial_\mu + i q A_\mu$,
where $q$ is the charge carried by the field upon
which the derivative operator acts. 
The total electromagnetic current is then
\begin{eqnarray}
J^\mu(x) &=& j^\mu_q(x) + j^\mu_\pi(x), \\
j^\mu_q(x) &=& \sum_a Q_a e \overline{q}_a(x) \gamma^\mu q_a(x),\label{quark}\\
j^\mu_\pi(x) &=& -i e [ \pi^\dagger(x) \partial^\mu \pi(x)
               -\pi(x) \partial^\mu \pi^\dagger(x)],\label{pion}
\end{eqnarray}
where $q_a(x)$ is the quark field operator 
for flavor $a$, $Q_a$ is its charge in units of e, and $e \equiv |e|$ 
is the magnitude of the electron charge.
The charged pion field operator,
$ \pi(x) = {1\over \sqrt{2}}[\pi_1(x) + i\pi_2(x)]$,
 either destroys a negatively charged pion 
or creates a positively charged one.

It is customary to define the helicity amplitudes for the electroproduction
of the delta as\cite{MUK87} 
\begin{eqnarray}
A_{3/2} &=& {1\over \sqrt{2 \omega_\gamma}}
\bra {\Delta; s_\Delta =3/2}  \vec{J}(0)\cdot \vec{\epsilon}
 \ket {N; s_N = 1/2}, \label{A3} \\
A_{1/2} &=& {1\over \sqrt{2 \omega_\gamma}}
\bra {\Delta; s_\Delta =1/2}  \vec{J}(0)\cdot \vec{\epsilon}
 \ket {N; s_N = -1/2}, \label{A1}
\end{eqnarray}
where the $\Delta$ is at rest and the photon is travelling along the z-axis
with 
%\begin{equation}
%q^\mu =( \omega, \vec{q}) , \,\,\, and \,\,\,  p^\mu_N = (E_N(q) , -\vec{q}).
%\end{equation}
%with $E_N(q) = \sqrt{M_N^2 + |\vec{q}|^2}$. 
right-handed 
polarization,  $\vec{\epsilon} = -{1\over\sqrt{2}} (1, i, 0)$.
The spin projections of $\Delta$ and $N$ 
along the z-axis are denoted as $s_\Delta$ and $s_N$ respectively.
%The energy of a real photon at the $\Delta$ resonance is
%$\omega_\gamma =  (M_\Delta^2 - M_N^2)/2M_\Delta$. 
For a virtual photon,
 the three-momentum in the $\Delta$ rest frame is given by 
\begin{equation}
|\vec{q}|^2 = Q^2 + {(M_\Delta^2 - M_N^2 - Q^2)^2 \over 4 M_\Delta^2},
\end{equation}
with $Q = \sqrt{-q^2}$ the magnitude of the four momentum transfer.
The photon energy is related to this by $q^2_0 = |\vec{q}|^2 - Q^2 $,
where for a real photon we have $Q^2 = 0$, so that 
$\omega_\gamma = |q_0| = (M_\Delta^2 - M_N^2)/2M_\Delta$.
%In the limit $Q^2 \rightarrow 0 $, this convention corresponds to the 
%usual helicity amplitudes for a real photon.
The experimentally extracted, resonant, helicity amplitudes 
are to be associated 
with the fully dressed initial and final baryons. 
In the cloudy bag model, due to the $\pi BB'$ coupling, a physical baryon
state is  described as a  mixture of a bare bag and its surrounding pion cloud,
\begin{equation}
\ket A = \sqrt{Z^A_2} [ 1 + (E_A - H_0 - \Lambda H_I \Lambda )^{-1} H_I ] 
\ket {A_0}, \label{state}
\end{equation}
where $Z^A_2$ is the bare baryon probability in the physical baryon states,
$\Lambda$ is a projection operator which projects out all the components
of $\ket A $ with at least one pion, and $H_I$ is the interaction Hamiltonian
which describes the process of emission and absorption of pions. 
The matrix element of $H_I$ between the bare baryon states 
is given by\cite{TT83,THESIS}
\begin{eqnarray}
v^{AB}_{0j}(\vec{k}) &\equiv& 
\bra {A_0} H_I \ket{{\bf \pi}_j(\vec{k}) B_0} \nonumber\\
 &=& {i f^{AB}_0\over m_\pi}
{u(kR) \over [2\omega_k (2\pi)^3]^{1/2}} \sum_{m,n}
C^{s_B m s_A}_{S_B 1 S_A} (\hat{s}^*_m \cdot {\vec k}) 
C^{t_B n t_A}_{T_B 1 T_A} (\hat{t}^*_n \cdot {\vec e}_j),
\end{eqnarray}
where the pion has momentum $\vec{k}$ and  isospin projection $j$,
 $f_0^{AB}$ is the reduced matrix element for the 
$\pi B_0 \rightarrow A_0$ transition vertex, $u(kR) = 3j_1(kR)/kR $,
$\omega_k = \sqrt{k^2+ m^2_\pi}$, and $\hat{s}_m$ and $\hat{t}_n$ 
are spherical unit vectors for spin and isospin, respectively.

\begin{figure}[t]
\begin{center}
%\rule{7cm}{0.2mm}
%\vskip 1.0cm
%\rule{7cm}{0.2mm}
\epsfig{figure=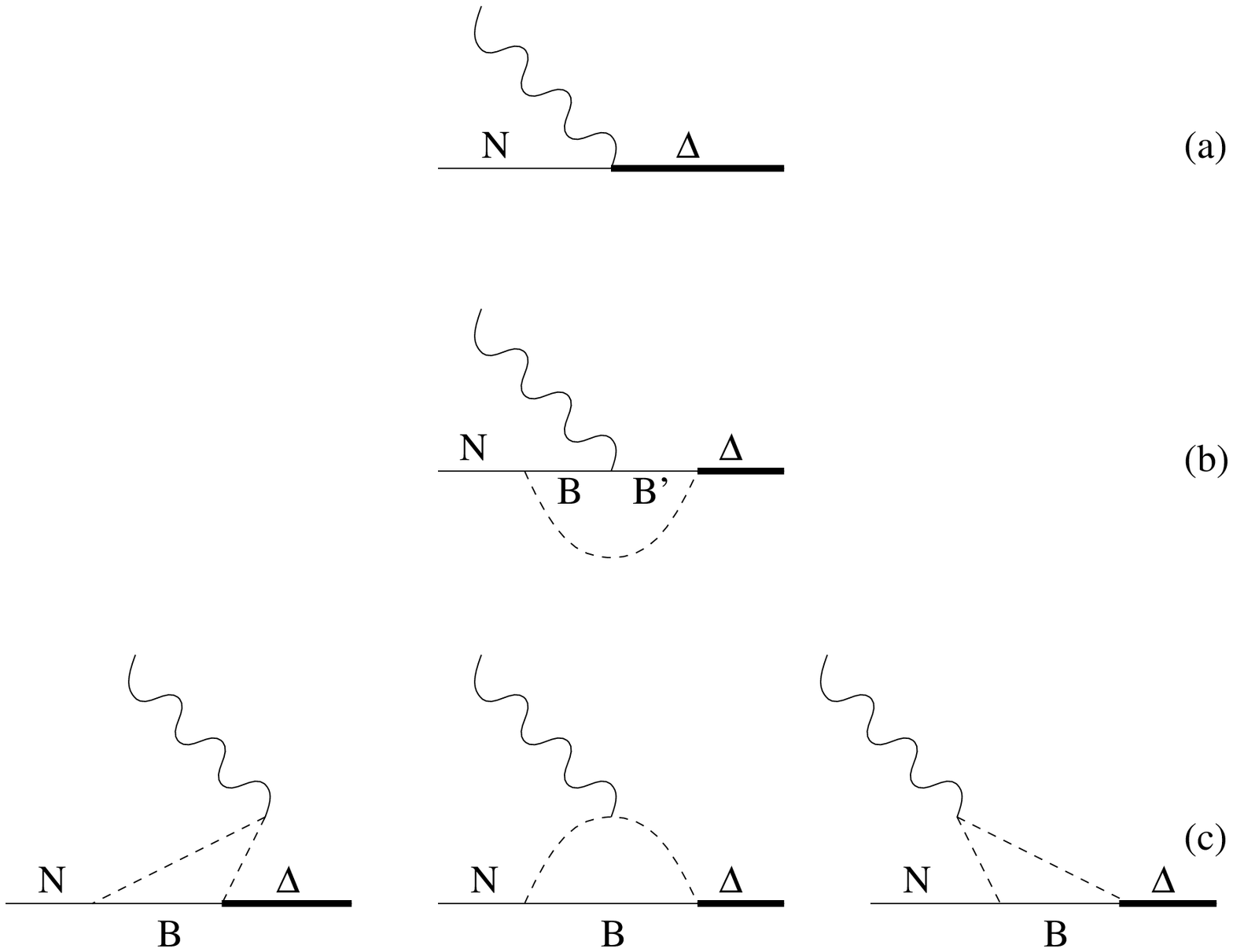,height=3.2in,width=4.2in}
%\def\epsfsize#1#2{0.5#1}
%\centerline{\epsfbox{F1.ps}}
\end{center}
\vskip 0.5cm
\fcaption{Diagrams illustrating the various contributions included 
in the calculation. The intermediate baryons $B$ and $B'$ are
restricted to the $N$ and $\Delta$ here.
\label{fig:radish}}
\end{figure}
%\begin{figure}
%\def\epsfsize#1#2{1.0#1}
%\centerline{\epsfbox{F1.ps}}
%\vspace{0.2in}
%\caption{ $N\rightarrow\Delta$ M1 transition form factor.}
%\twospace
%\label{flatt}
%\end{figure}

Under the approximation  that there is
at most one pion in the air, there are three different
processes contributing to the $\gamma^* N \rightarrow\Delta$ vertex, 
as shown in Fig.~1.
For Figs.~1(a) and 1(b), we substitute Eqs.~(\ref{quark}) 
and (\ref{state}) into
Eqs.~(\ref{A3}) and (\ref{A1}), and obtain the helicity  amplitudes,
%corresponding to Figs. 1(a) and 1(b),  we obtain
%Using the above defintions and  SU(6) wave functions for the spin-flavor part
%of baryon wave functions, 
%obtain, for Figs. 1(a) and 1(b),
%
\begin{eqnarray}
A^{(a)}_{3/2}(Q^2) &=& \sqrt{3} A^{(a)}_{1/2}(Q^2) =  
A_{\rm{bare}}(Q^2) \sqrt{Z^N_2 Z^\Delta_2}, \\
A^{(b)}_{3/2}(Q^2) &=& \sqrt{3} A^{(b)}_{1/2}(Q^2) = 
A_{\rm{bare}}(Q^2) {(f^{NN})^2 \over 27\pi^2m_\pi^2} 
\int\! {dk k^4 u^2(kR)\over \omega_k} \left[
  {5/4 \over \omega_k (\omega_k + \delta - \omega_\gamma)}\right. \nonumber \\
&&\left.+ {1 \over (\omega_k + \delta)(\omega_k + \delta - \omega _\gamma)}
+ {2/25 \over (\omega_k + \delta)(\omega_k - \omega_\gamma)}
+ {1 \over \omega_k (\omega_k - \omega_\gamma)} \right], \label{AB}
\end{eqnarray}
where $\delta = m_\Delta - m_N$, and  $f^{NN}$ is the renormalized $\pi NN$
coupling constant. 
The four terms in the right-hand side of Eq.~(\ref{AB})
correspond to four possible intermediate states, $(N\Delta)$, $(\Delta\Delta)$,
$(\Delta N)$, and $(N N)$, respectively. 
The recoil corrected bare $\gamma N_0\rightarrow\Delta_0$ 
transition amplitude is
\begin{equation}
A_{\rm{bare}}(Q^2) = -{ e\over \pi\sqrt{6\omega_\gamma}} 
{\int_0^R\! dr r^2 g(r) f(r) j_1(qr) K(r)
\over \int_0^R\! dr r^2 \rho(r) K(r)},
\end{equation}
where $K(r) = \int\! d^3x \rho(\vec{x}) \rho(-\vec{x} - \vec{r})$
 is the recoil function to account for the correlation of the 
two spectator quarks. 
The renormalization constants, $Z^A_2$, are  determined by the normalization
condition for the physical baryon state\cite{TT83}.
%, i.e. 
In this work, we have adopted the usual philosophy for the renormalization
in the CBM. Throughout this work approximate relation,
$ f^{AB} \simeq \left({f^{AB}_0\over f^{NN}_0}\right) f^{NN}$, is always used.
There are uncertain corrections on the bare coupling contant $f^{NN}_0$,
such as the nonzero quark mass and correction of center of mass motion.
Therefore, we use the renormalized coupling constant in our
calculation, $f^{NN} \simeq 3.03$, 
which correspond to the usual $\pi NN$ coupling constant, 
$f^2_{\pi NN}\simeq 0.081$. 
As a result, the factor $\sqrt{Z^N_2 Z^\Delta_2}$ is absorbed into the 
renormalized coupling constants in Fig.~1(b). This treatment is equivalent
to the original CBM formalism
up to order $(f^{NN})^2$ and consistent with current conservation.

To evaluate the contribution caused by 
the photon-pion-pion coupling vertex [see Fig.~1(c)],
we use the usual plane wave expansion for the quantized pion field
\begin{equation}
\pi_j(\vec{x},t=0) = \int\! {d^3k\over [(2\pi)^3 2\omega_k]^{1/2}}
\left[ a_j(\vec{k}) e^{i\vec{k}\cdot\vec{x}} + 
a^\dagger_j(\vec{k}) e^{-i\vec{k}\cdot\vec{x}}\right],
\end{equation}
where $a_j(\vec{k})$  $(a^\dagger_j(\vec{k}))$ annihilates (creates) a pion
with momentum $\vec{k}$ and isospin $j$. 
With the identity,
$a_j(\vec{k})\ket A = (E_A - \omega_k - H)^{-1} H^\dagger_{I}(\vec{k},j)
\ket A, $ 
%\\
%a_{j'}(\vec{k'}) a_j(\vec{k})\ket A &=& 
%(E_A - \omega_k - \omega_{k'} - H)^{-1} V^\dagger_{0j}(\vec{k})
%(E_A - \omega_{k'} - H)^{-1} V^\dagger_{0j'}(\vec{k'}) \nonumber \\
%&+& 
%(E_A - \omega_k - \omega_{k'} - H)^{-1} V^\dagger_{0j'}(\vec{k'})
%(E_A - \omega_{k} - H)^{-1} V^\dagger_{0j}(\vec{k}) \ket A,
%\end{eqnarray}
we obtain the transition amplitude at position $\vec{x}$,
\begin{eqnarray}
\bra{\Delta,s_\Delta} \vec{j}_\pi(\vec{x}) \ket{N,s_N} 
&=& -i e \sum_{j j'}\epsilon_{jj' 3} \int\! d^3k d^3k'
e^{i (\vec{k} - \vec{k'})\cdot \vec{x}} 
{\vec{k} \over (2\pi)^3 2(\omega_k\omega_{k'})^{1/2}} \nonumber \\
&\times&
\sum_B \left[ \eta^B_{j'j}(\vec{k'},\vec{k}) G^B(\vec{k'},\vec{k}) +
               \eta^B_{jj'}(\vec{k},\vec{k'}) G^B(\vec{k},\vec{k'})\right].
\end{eqnarray}
Here $B$ denotes the intermediate baryon states 
(restricted to $N$ and $\Delta$ here), and 
\begin{eqnarray}
\eta^B_{j'j}(\vec{k'},\vec{k}) &\equiv& {f^{\Delta B} f^{NB} \over m^2_\pi}
{u(kR) u(k'R) \over 16\pi^3 (\omega_k \omega_{k'})^{1/2}}
 \sum_{s_B} 
 C^{s_B m' s_\Delta}_{S_B 1 S_\Delta} C^{s_B m s_N}_{S_B 1 S_N}
             (\hat{s}^*_{m'} \cdot {\bf k'}) (\hat{s}^*_{m} \cdot {\bf k})
\nonumber \\
&\times& \sum_{t_B}
C^{t_B n' t_\Delta}_{T_B 1 T_\Delta} C^{t_B n t_N}_{T_B 1 T_N}
             (\hat{t}^*_{n'} \cdot {\bf e}_j)(\hat{t}^*_{n} \cdot {\bf e}_j),\\
%%%%%
G^N(\vec{k'},\vec{k}) &\equiv& 
  {1\over (\omega_k + \omega_{k'} + \delta)\omega_{k}}
+ {1\over (\omega_{k'} -\omega_\gamma)\omega_{k}}
+ {1\over (\omega_{k'} -\omega_\gamma)(\omega_k + \omega_{k'} -\omega_\gamma)},
\label{gn} \\
G^{\Delta}(\vec{k'},\vec{k}) &\equiv&
  {1\over (\omega_k + \omega_{k'} + \delta)(\omega_k + \delta)}
+ {1\over (\omega_{k'} + \delta - \omega_\gamma)(\omega_k + \delta)}\nonumber\\
&&
+ {1\over (\omega_{k'} + \delta - \omega_\gamma)
 (\omega_k + \omega_{k'} - \omega_\gamma)}. \label{gd}
\end{eqnarray}
$G^{\Delta}(\vec{k},\vec{k'})$ and $G^{\Delta}(\vec{k},\vec{k'})$ are obtained
by the interchange of $\vec{k}$ and $\vec{k'}$ in the corresponding equation. 
 The three terms in Eqs.~(\ref{gn}) and (\ref{gd}) indicate the three
different time orders in the time-ordered perturbation theory, as illustrated
in Fig.~1(c).
%The contribution to the $\gamma^* N\rightarrow\Delta$  helicity amplitudes 
%are thus obtained approximately through a Fourier transform. This leads to 
Using the translational invariance of the electromagnetic current operator,
$j^\mu(x) = e^{i\hat{p}\cdot x} j^\mu(0) e^{-i\hat{p}\cdot x}$, then
the $\gamma^* N\rightarrow\Delta$  helicity amplitudes due to the 
$\gamma \pi\pi$ interaction are simply given by 
\begin{equation}
A(Q^2) = \int\! d^3x e^{i \vec{q}\cdot\vec{x}}
\bra{\Delta,s_\Delta} \vec{j}_\pi(\vec{x})\cdot \vec{\epsilon}\ket{N,s_N}.
\end{equation}
After performing some  spin and isospin  algebra, we obtain
\begin{eqnarray}
A^{(c)}_{3/2}(Q^2) &=&  
- {(f^{NN})^2 |\vec{q}| \over 240\sqrt{6\omega_\gamma}\pi^3 m_\pi^2} 
\int\! {d^3k k^4 \sin^2\!\theta \, u(kR) u(k'R)\over  \omega_k \omega_{k'}}
\nonumber\\
&\times& \left[ G^N(\vec{k}, \vec{k'}) + 
3 G^\Delta(\vec{k'}, \vec{k}) + 2 G^\Delta(\vec{k},\vec{k'} )\right], \\
A^{(c)}_{1/2}(Q^2) &=&  
- {(f^{NN})^2 |\vec{q}| \over 720\sqrt{2\omega_\gamma}\pi^3 m_\pi^2} 
\int\! {d^3k k^4 \sin^2\!\theta \, u(kR) u(k'R)\over  \omega_k \omega_{k'}}
\nonumber\\
&\times& \left[ 2 G^N(\vec{k'}, \vec{k}) - G^N(\vec{k}, \vec{k'}) + 
 G^\Delta(\vec{k'}, \vec{k}) + 4 G^\Delta(\vec{k},\vec{k'} )\right],
\end{eqnarray}
where 
$\vec{k'} = \vec{k} + \vec{q}$, $\omega_{k'} = \sqrt{k'^2 + m^2_\pi}$,
and $\theta$ denotes the angle between $\vec{k}$ and $\vec{q}$.
It is worthwhile to  mention that the form of our results for Fig.~1(c) 
are quite different
from those of KE\cite{KE83} and Bermuth et al.\cite{Bermuth86} where
the integral variables are $k$ and $k'$ in their formulations. We believe that 
our expressions are more straightforward and manifestly  
respect the three momentum conservation
at the $\gamma\pi\pi$ vertex.
%The support for the $\theta$ integral is relatively simple,
%is another advantage of our form as far as the numerical evaluation concerns. 

\section{Results}
\begin{figure}[t]
\begin{center}
\epsfig{figure=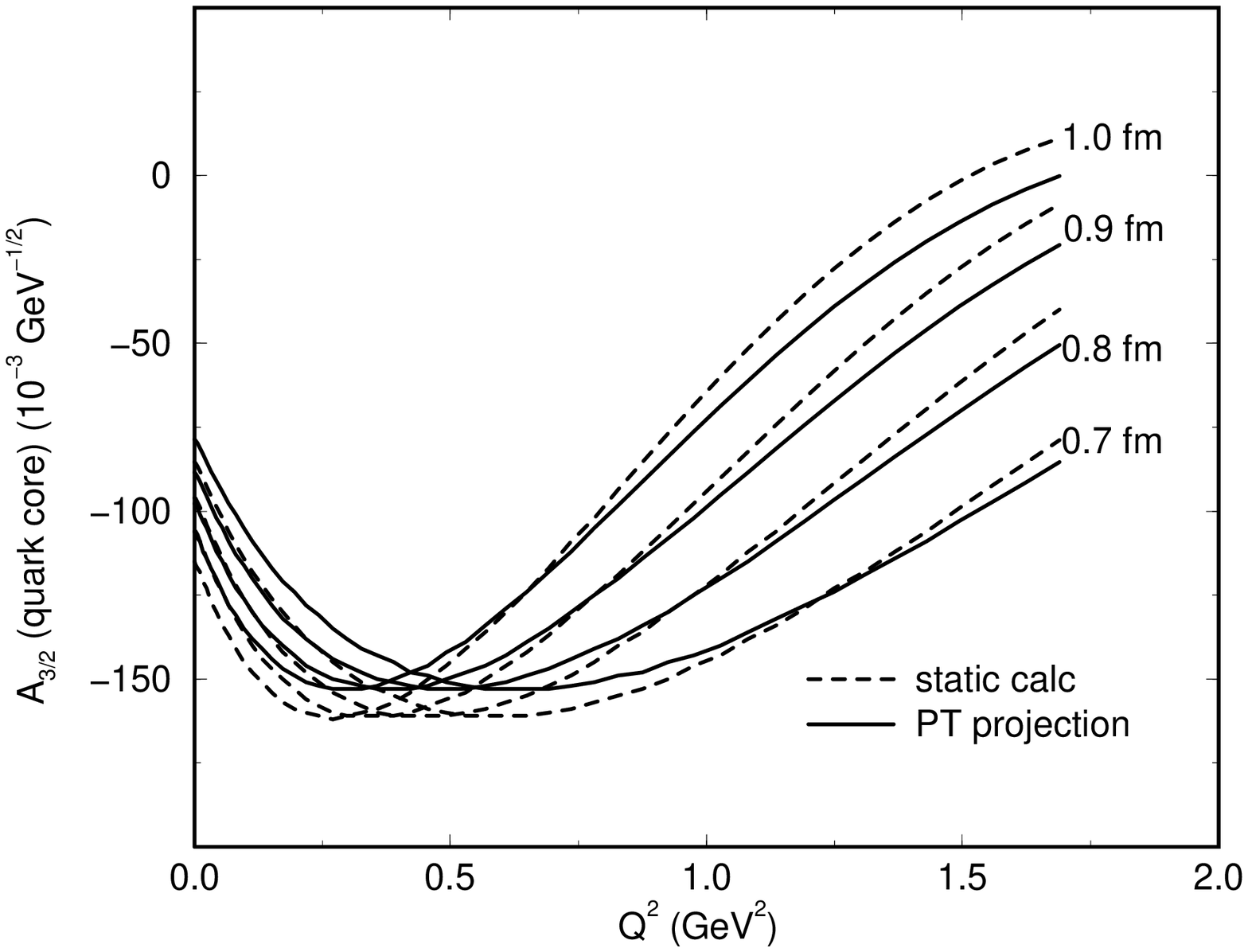,height=3.2in,width=4.2in}
\end{center}
\fcaption{The effect of the center of mass correction on the 
helicity amplitude, $A_{3/2}$, for the bare bag. The number 
on each curve indicates the bag radius in fm  for the calculation.
\label{fig:bare}}
\end{figure}
%Figure 2 shows the recoil function $K(r)$ due to the two quark spectators. 
%It cuts down  the contributions from
%the quark wave functions near the bag boundary by  up to 50 \%. 
The overall effect of the PT recoil
correction on the bare bag contribution to the 
typical $\gamma^* N\rightarrow\Delta$ helicity amplitude,
 $A_{3/2}$, 
is shown in Fig.~\ref{fig:bare}. 
In the real photon limit ($Q^2 \rightarrow 0$), 
the magnitude of the $\gamma^* N\rightarrow\Delta$ transition
amplitude increases with the bag radii in a fashion similar to 
that of the magnetic moment
of bare baryons. The correction of the center of mass motion usually reduces
 the bare transition
amplitudes by 5 to 10 \% for $Q^2 \alt 0.5 \mbox{ GeV}^2$ within a reasonable
range of bag radii. However, this recoil correction would flip sign and
 increase the transition amplitude for larger momentum transfers.

%\begin{figure}[t]
%\begin{center}
%\epsfig{figure=fig3.ps,height=3.4in,width=5.0in}
%\end{center}
%\fcaption{The recoil function K(r), defined following Eq.\ (23),
% due to the two quark spectators in  arbitrary units.
%\label{fig:fig.rec}}
%\end{figure}

%\begin{figure}[t]
%\begin{center}
%\epsfig{figure=F3.ps,height=3.4in,width=4.5in}
%\end{center}
%\fcaption{The effect of the center of mass correction on the 
%helicity amplitude, $A_{3/2}$, for the bare bag. The number 
%on each curve indicates the bag radius in fm  for the calculation.
%\label{fig:bare}}
%\end{figure}

%reverse the sign of the
% correction for larger momentum transfers
%for larger momentum transfers.
%It is interesting to see that for $Q^2 \geq 0.5 \mbox{ GeV}^2$, the helicity
%amplitude calculated from the bare bag decreases with 
%the bag radius.

\begin{figure}[t]
\begin{center}
\epsfig{figure=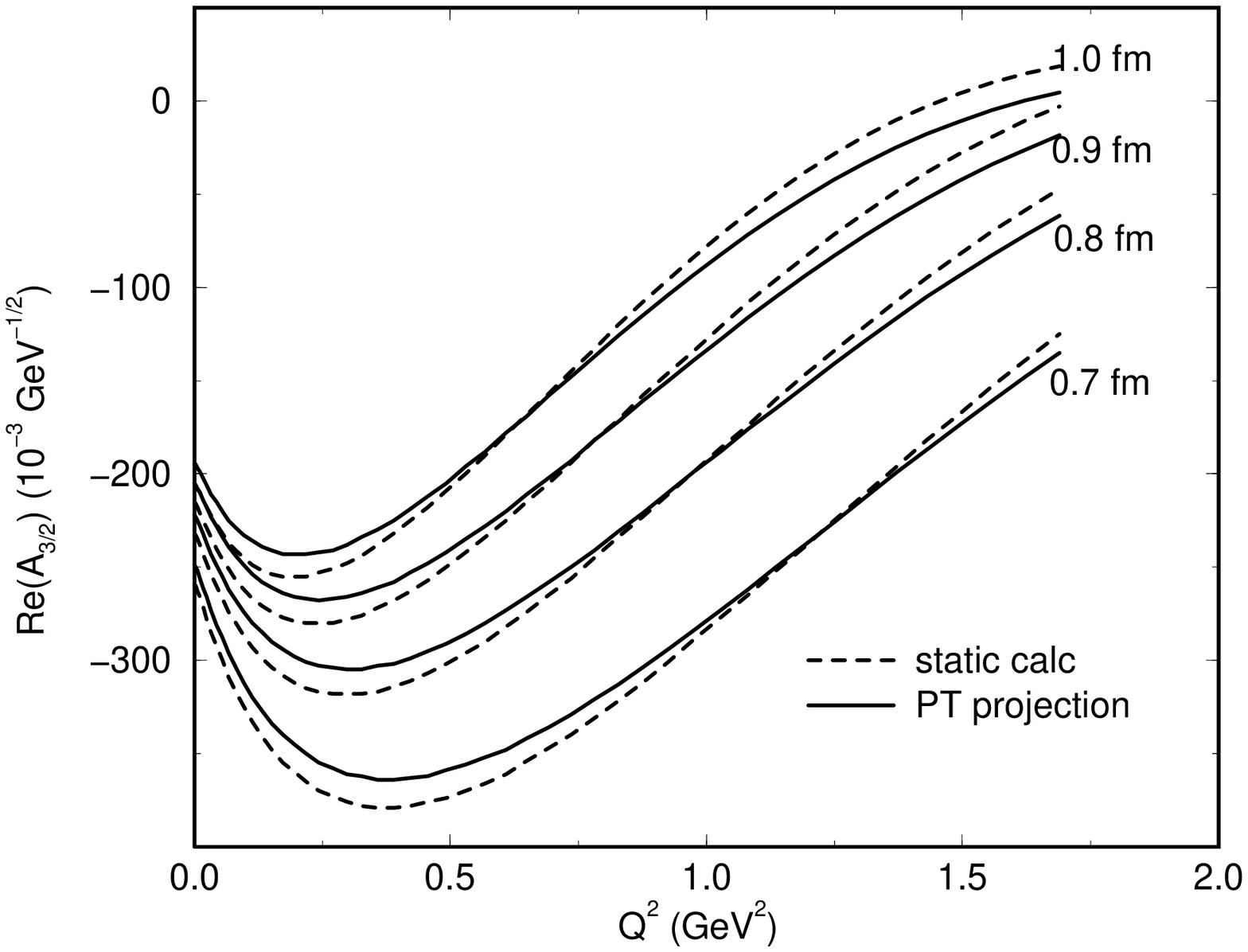,height=3.2in,width=4.2in}
\end{center}
\fcaption{The real parts 
of total $\gamma^*N\rightarrow\Delta$ helicity amplitudes, 
Re[$A_{3/2}(Q^2)$].
\label{fig:helicity}}
\end{figure}

The real parts of total helicity amplitudes, 
$A = A^{(a)} + A^{(b)} + A^{(c)}$, 
%with four different bag radii 
are presented
in Fig.~\ref{fig:helicity}. With the contributions of the pion cloud,
the bag radius dependence is quite different from that for the bare transition
amplitudes shown in Fig.~\ref{fig:bare}. This can be explained
by the fact that 
the pionic contribution is competing with that of  quark core,
since a small bag radius means a strong pion cloud. 
In the small $Q^2$ region, the pion cloud compensates more 
than the loss in the bare amplitude when using a 
small bag radius. 
Generally, the smaller the bag radius, 
the larger the total transition amplitude.
We list the helicity amplitudes corresponding to the  real photon limit
at the $\Delta$ resonance  in Table 1.
%Notice that, since the $\Delta$ is a resonance, 
%the helicity amplitudes are actually complex.
%In the CBM the imaginary part arises from  thresholds in the 
%pion loop integral.  
With the small bag radius  $R$ = 0.7 fm, 
we are able to reproduce the experimental helicity
amplitude in this model.

%\begin{figure}[t]
%\begin{center}
%\epsfig{figure=F5.ps,height=3.4in,width=4.5in}
%\end{center}
%\fcaption{The real parts 
%of total $\gamma^*N\rightarrow\Delta$ helicity amplitudes, 
%Re[$A_{3/2}(Q^2)$].
%\label{fig:helicity}}
%\end{figure}

\begin{table}[t]
\protect
\tcaption{Helicity amplitude of delta photoproduction, $A_{3/2}$, 
in units of $10^{-3} \mbox{GeV}^{-1/2}$.
Here static denotes the static calculation and PT denotes the Peierls-Thouless
projection. The indices a, b, and c 
correspond to the Figs.~1(a), 1(b), and 1(c) respectively. 
The latest estimate by Particle Data Group is $-258 \pm 6$. }
%\small
\vspace{0.4cm}
\begin{center}
\begin{tabular}{l|cccc|cccr}
%\hline\hline
 & \multicolumn{4}{c|}{static}   &  \multicolumn{4}{c}{PT}  \\ \hline
  R(fm)& a & b & c & total & a & b & c & total \\ \hline
 1.0 &-115 & -53 -21i & -64 -18i & -205 &-106 & -49 -19i & -64 -18i & -195\\
 0.9 &-106 & -62 -20i & -80 -20i & -216 & -98 & -58 -18i & -80 -20i & -205\\
 0.8 & -96 & -73 -19i &-101 -21i & -233 & -86 & -68 -17i &-101 -21i & -222\\
 0.7 & -85 & -87 -17i &-129 -22i & -260 & -79 & -80 -16i &-129 -22i & -249
%\hline
\end{tabular}
\end{center}
\end{table}

\section{Summary}
In conclusion, we have calculated the 
$\gamma^* N \rightarrow\Delta$ transition 
form factors in the cloudy bag model, 
including the center of mass correction via 
a simplified Peierls-Thouless projection method.
The effect of this recoil correction is to slightly 
reduce the magnetic form factor
at small momentum transfer and to  
make the form factor slightly  harder. Generally, with the PT projection
the transition moment is reduced by about $5 \sim 8 \%$.

The pion cloud contribution proved to be crucial to account for 
the measured helicity amplitudes using a reasonable bag radius in this model.
In similar calculations using constituent quark models 
(the nonrelativistic\cite{IK82} and relativized quark models\cite{WARNS90}), 
 the helicity amplitudes
are usually significantly underpredicted. 
Further details and extensions of this work will be presented 
elsewhere\cite{delta96}.

This work was supported by the Australian Research Council.

\end{document}